\documentclass[a4paper,12pt,twoside]{report}
\usepackage[left=2cm,right=2cm,top=2cm,bottom=3cm]{geometry}

\usepackage{ifthen} 
\newboolean{pdflatex}
\setboolean{pdflatex}{true}

\newboolean{uprightparticles}
\setboolean{uprightparticles}{false} 

\newboolean{inbibliography}
\setboolean{inbibliography}{false} 

\usepackage{subcaption}
\usepackage{makecell}

\def\mm{\ensuremath{\mu\mu}\xspace}

\def\invRJPsi{\ensuremath{r_{\jpsi}^{-1}}\xspace}

\def\invRpK{\ensuremath{R_{\proton\kaon}^{-1}}\xspace}

\def\LbTopKee{\decay{\Lb}{p\kaon \epem}}

\def\LbTopKee{\decay{\Lb}{\proton\kaon \epem}}

\usepackage{longtable} 

\usepackage{graphicx}
\usepackage{verbatim}
\usepackage{latexsym}
\usepackage{mathchars}
\usepackage{setspace}

\input{blocked.sty}
\input{uhead.sty}
\input{boxit.sty}
\input{icthesis.sty}

\newcommand{\NN}{{\sf I\kern-0.14emN}}   
\newcommand{\ZZ}{{\sf Z\kern-0.45emZ}}   
\newcommand{\QQQ}{{\sf C\kern-0.48emQ}}   
\newcommand{\RR}{{\sf I\kern-0.14emR}}   

\newcommand{\diff}[1]{\mathrm{\frac{d}{d\mathit{#1}}}}

\newcommand{\normallinespacing}{\renewcommand{\baselinestretch}{1.5} \normalsize}

\newcommand{\syncc}{~\stackrel{\textstyle \rhd\kern-0.57em\lhd}{\scriptstyle L}~}

\begin{document}

\tableofcontents
\cleardoublepage

\title{\LARGE {\bf Would you rather be a fish ? }\\
 \vspace*{6mm}
}

\author{Yasmine Sara Amhis}
\submitdate{May 2020}

\normallinespacing
\maketitle

\preface

\begin{abstract}

This manuscript summarises a selection of my research activities from the past ten years. I obtained my PhD in 2009 at  LAL  under the supervision of Marie-H\'el\`ene Schune and Jacques Lefran\c{c}ois.  Afterwards I moved to EPFL as a postdoctoral researcher, where I worked for three years in the group of Olivier Schneider. 
In 2012, I obtained a CNRS position at LAL now IJCLab. 
Three physics analyses based on the LHCb data, focusing on properties of \bquark-baryons such as masses, lifetimes and a lepton universality test are discussed in this manuscript. For each topic, I cherry picked key experimental points that will be  developed in the text. The papers corresponding to these analyses can be found at the end of the document.  Specific contributions to the scintillator fibre tracker and the calorimeters of the LHCb detector are also described.  The object that  you are holding in your hands is not meant to be pedagogical nor an exhaustive review, however, I hope that it contains enough information and references to guide any curiosity that it triggers. 

\end{abstract}

\cleardoublepage

\addcontentsline{toc}{chapter}{Acknowledgements}

\begin{acknowledgements}


\small
\begin{flushright}
Bagneux, March 20$^{th}~2020$\\
{\it Day 4 of confinement}
\end{flushright}

I will be honest, I am writing these acknowledgements standing in my kitchen, while Loulou is watching {\it ``Youhoo to the rescue"}. What a strange situation we find ourselves in. Nonetheless, I would like to use this precious time to acknowledge a few people.

I would like to first thank my referees, Tulika Bose, Guy Wilkinson and Isabelle Wingerter-Seez for their comments and suggestions to improve this manuscript. I would like to also thank Gino Isidori, François Lediberder and Achille Stocchi for accepting to be in my habilitation Jury.

 Renato Quagliagni for his robustness like a good pattern recognition algorithm, with whom I had a great time figuring out the tricks of the LHCb tracking and for investigating the properties of all the hits in the {\it SciFi} one by one. 

Vitalii Lisovskyi for his mythical "maybe", for his curiosity, intelligence,  enthusiasm and ability to fish out every last single little bias and background in the data.   

Carla Marin Benito, for ``guarding the house" while I was busy writing, for being sharp like a Japanese Samurai sword and for being such a joy to work with.

I would like to thank Anja Beck for making the helicity formalism sound like a piece of vegan cake.

I would like to thank Marie-Hélène Schune for being the first one to open the door of LAL and building 208 in particular to me. I am looking forward to many H$_{2}$O discussions.

I would like to thank Jacques Lefrançois, we should all want to be Jacques when we grow up, the world would be a much better place.

I appreciated during these years the interactions with all the PhD students in the LHCb group, and I would like to thank them all. Alexandra Martín Sánchez with whom I ended up sharing one of our bigger life experiments, Olga Kochebina for her good spirits and knowledge about charm physics, Maksym Teklishyn my office mate for a year,  Alexis Vallier who's good mood is difficult to challenge, Martino Borsato and his ``interesting" and very specialised taste in music, Victor Daussy-Renaudin for recommending all the podcasts popular to trendy left wing kids, Andrii Usachov for his weird stories,  Fabrice Desse for his unbeatable cool, and Elisabeth Niel for making me smile even when I am grumpy and also for being so precise with $^{\bf 6}_{\bf 8}$ rythms. %

I would like to thank Francesco Bossu and Michael Winn, for putting in a new light the heavy ions topic, that was not an easy task. 
 
 Sergey Barsuk for his odd ukrainian chocolate imports and upgrading our group with a decent coffee machine.
 
 Patrick Robbe for his friendship, knowledge about the of the gory details of the LHCb simulation, as well as harmony rules.

 I would like to thank Olivier Schneider, Aurelio Bay, Tatsuya Nakada, Raluca Muresan, Marc-Olivier Bettler, G\'eraldine Conti and Joel Bressieux for the wonderful three years at EPFL.

I would like to thank my theory colleagues, Sébastien Descotes-Genon for the hours that he spared me to discuss EFTs, Damir Bečirević, for his invaluable help and also for quizzing me like a student about the dimensions of operators. I would like to thank Olcyr Sumensari for proofreading the theory chapter and very precise comments on Figure 1.3. I would also like to thank Danny van Dyk, Marzia Bordone,  Alexandre Lenz, Aoife Bharucha, Peter Stangl, Diego Guadagnoli, and Yossi Nir for all the discussions. 

A special thank you to Tim Gershon, who's always been so very supportive throughout all the years, for his help reviewing all my relevant documents. Thank you. 

Peter Clarke, for the great time we had as conveners and discussing the lifetime agonies. 

I would like to thank my collaborators of the lifetime and mass measurements Francesca Dordei, Sneha Malde, Greig Cowan and Matthew Needham. Jaap Panman for the luminosity work, Pierre Billoir and Manuel Schiller for the seeding discussions. Mark Tobin for his help during the ST piquets.

Sophie Redford for always correcting my English, Tim Head for teaching us about PR and MR and all of that. Conor Fitzpatrick for his general Irishness and everything that goes with it. Ben Couturier for his help with the obscure part of the LHCb software. Evelina Gersabeck for her support. Eluned Smith for her initiative and cello chats, Paula Collins for building us a Velo. Simone Bifani for his help swinging an epic EW talk, Stephanie Reichert for her  feminism and careful readings. The Oxbridge representatives of LHCb, Patrick Owen and Matthew Kenzie for the long debates we had about the anomalies and other topics.  Julian Wishahi for his support during the early SciFi days.   
Jerko Merkel for the fun  chats at CERN and attempts to teach me German. 

My (loud) colleagues from building 208 and in particular Nicolas Leroy for the book recommendations. Laurent Serin for sharing the L2 class with me and Antoine Laudrain for his help preparing the NPAC projects. Dimitris Varouchas for his Greekness, Marta Spinelli for her energy. 

Francesco Polci, for his kindness, for being a  great collaborator and friend, thank you for the teas in R1, the music and  all the  complaining. 

Viola Sordini, for sharing so many chapters of our lives together, for offering me her couch when I moved to Geneva, for the hours and hours of talking and everything else. 

Zaida Conesa del Valle for being such a great collaborator throughout our various committees.  

I would like to thank Nathalie Grub, Lalaine Barbon Strebel, Anu Liisa Saarelainen, Cindy Denis, Amélie Caillet, Erika Luthi, Isabelle Vauleon, Dominique Bony, Catherine Bourge, Catherine Zomer, and Sylvie Teulet for making my administrative duties an order of magnitudes easier. 

Gregory Perrin, Thomas Roulet, Guillaume Philippon, Gerard Marchal Duval for their help setting up the masterclasses every year. 

Outside of the work environment I benefit from a very complex and sophisticated architecture of humans and I would like to thank them all. 

Mathilde Daussy-Renaudin, for being the best supportive petite chauve-souris and for her infinite knowledge about medieval witches and snails. 

Emmanuelle Zagoria, who helped me outline so many projects and hopefully many others to come,  for being such an inspiration and for somehow creating a little bit of New York just for me.  

Blandine Huchet for making me discover so many  string quartets and the time spent in chamber music festivals. 

 My cello teachers, Guillaume Martign\'e, Domitille Sanyas, et Thalie Bonvallet Michalakakos who have been through their teachings, the guardians of my sanity and curiosity.

Ljubica and Vladimir for their kindness and especially for looking after Loulou, taking such great care of her during all the hours when I am working or travelling. 

Julien Bouyssou et Aghiles Ziani, who stacked one on top of the other would reach 4m. Thank you for being the brothers I never had. 

My sisters, Lili et Nina, for their magnificent sense of humour and for their million jokes in all circumstances, and by all circumstances I mean all circumstances, you two are the pillars of my joy.

My parents, for the eternal love and support, their unshakable confidence,  where would I be without you? Most probably nowhere. 

My grand mother for being the first one to teach me about patriarchy and how we should slowly and methodically dismantle it, thank you for  being so stubborn and an inspiration for finishing your eleventh book at age ninety two.

Finally, I would like to  thank my ultimate partner in crime.  I could make things easy, but then again, why would I? 

\begin{centering}
{*****}
\end{centering}
\newpage 
\subsection*{Audio support}
While writing this document, or driving in the car thinking about it,  I listened to: 
\begin{itemize}
    \item Johannes Sebastien Bach - Goldberg variations and Cello Suites. 
    \item Anoushka  Shankar -  Land of gold. 
    \item System of a Down -  Toxicity. 
    \item Britney Spears - Everything. 
    \item Antonio Vivaldi -  Double Cello Concerto 531. 
    \item Jeanne Added  - Radiate. 
    \item Clara Lucia - La grenade. 
    \end{itemize}
for many, many hours. 
\end{acknowledgements}
\cleardoublepage


\newpage

\begin{figure}[h]
\centering
\includegraphics[width=0.8\textwidth]{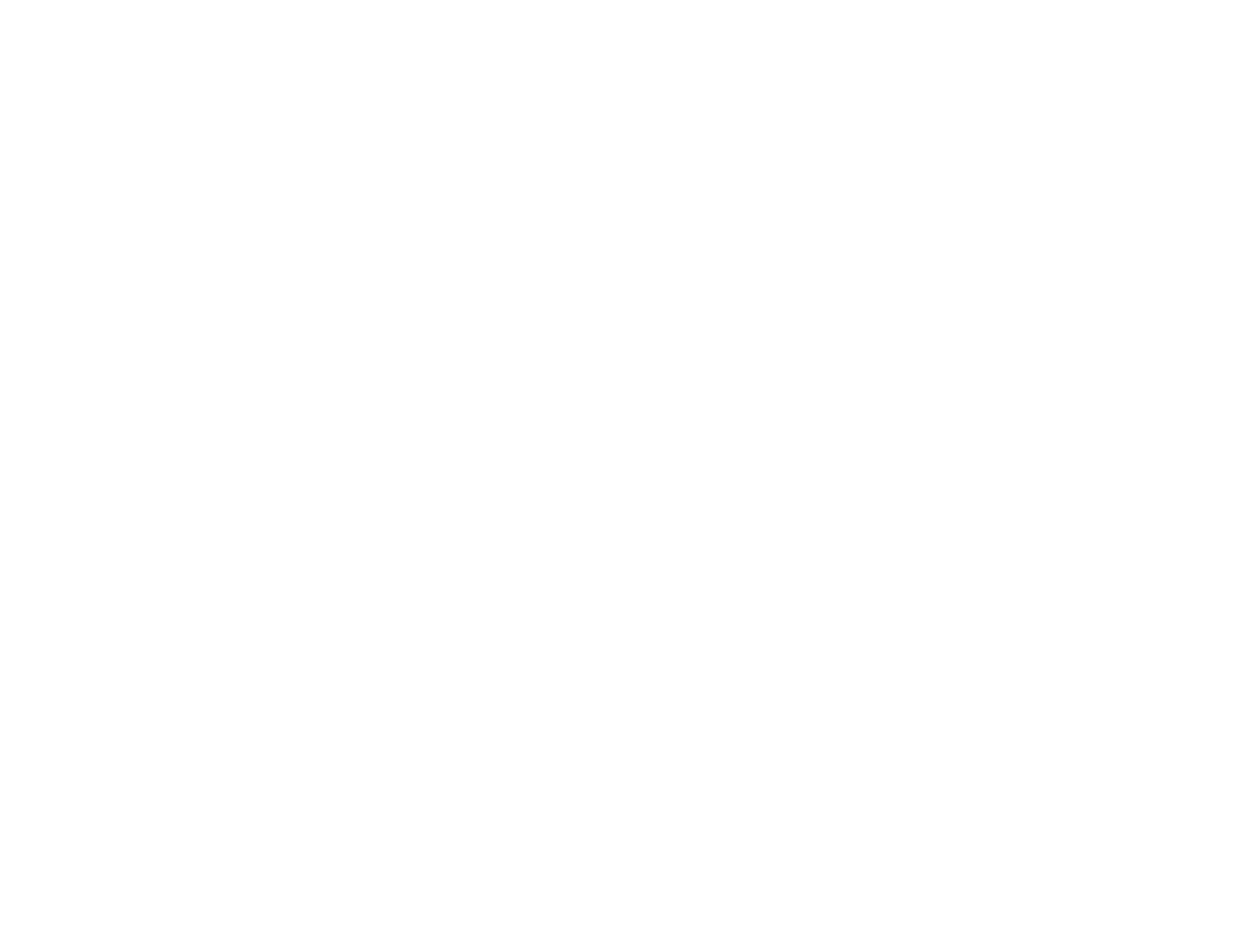}
\end{figure}

\begin{flushright}
\textrm {\it Pour un kilo de Loulou}
\vfil
\end{flushright}
\clearpage
\input{quotes/quotes}

\body
\input{introduction/introduction}
\input{background/background}


\chapter{Conclusion}
\label{ch:conclusions}

\epigraph{\textit {``Que mon conte soit beau et se d\'eroule comme un long fil"}}{Marie Louise Taous Amrouche
Le grain magique}

I had the good fortune to contribute to a few, yet very diverse facets  of flavour physics, as well to the LHCb detector and its first upgrade.  The months ahead of us are extremely important, as we are about to  embark   in the commissioning of an almost whole new detector, while working on the analysis of the legacy datasets and preparing LHCb upgrade II. Truly  somedays, it  seems like a titanic work. However, the excitement and the myriad of possibilities from flavour physics are such, that I believe all the efforts are worth it (although to be honest, sometimes after sitting in four or five hours long meetings, discussing the impact of the phases of the moon on our measurements of detector designs,  I do think that all of it should be tossed out a window). 

The results presented in this habilitation explored different properties of \bquark-baryons. 
To achieve  precise mass measurements of the \Lb, \Xib and \Omegab baryons, it was shown that a good control and understanding of the momentum scale played an essential role. The \Omegab measurement allowed to clarify the observed discrepancy between the two previous measurements from the Tevatron.  I then proceeded to discuss lifetime measurements of the \bquark-hadrons using decay modes with \jpsi in the final state. Given the size of the statistical uncertainties, it was mandatory to have an excellent handle on the systematic uncertainties.  With this analysis, lifetime acceptance effects were broken down as much as physically possible. Data-driven methods were employed to estimate the impact of reconstruction effects.  The results were in good agreement with the theoretical prediction from HQET and contributed to significantly  reduce the uncertainty of the updated world averages.
Finally, I discussed the first test of lepton universality using \Lb decays, \invRpK. With this analysis, the first observation of the rare mode \LbTopKee was established as well as the measurement of the branching fraction of  \LbTopKee  in $\qsq \in [1,6]$ \gevgevcccc and $m_{pK} <$ 2600 \mevcc. A key aspect of this analysis relies on the computation of the observable  \invRJPsi which gives confidence in the good control of the reconstruction effects and expected differences between electrons and muons.   While this is statistically limited, it follows the same trend observed in  lepton universality tests performed with other $B$-mesons.  With this  measurement, we did not close the window on the anomalies, so  let us see if they will remain and lead to the discovery of new particles, or fade away with more data. 

To conclude, the experimental measurements of \bquark-baryon properties contributed to the understanding or validation of theoretical computations.  However, there are still plenty of measurements to be done and phenomena to be further investigated, and I am looking forward to doing so in LHCb in the next few years.

\appendix
\addcontentsline{toc}{section}{References}
\setboolean{inbibliography}{true}
\bibliographystyle{LHCb}


\begin{thebibliography}{10}

\bibitem{PDG}
M.~Tanabashi et~al.
\newblock {Review of Particle Physics}.
\newblock {\em Phys. Rev.}, D98(3):030001, 2018.

\bibitem{Planck}
P.~A.~R. Ade et~al.
\newblock {Planck 2015 results. XIII. Cosmological parameters}.
\newblock {\em Astron. Astrophys.}, 594:A13, 2016.

\bibitem{Higgs}
Georges Aad et~al.
\newblock {Measurements of the Higgs boson production and decay rates and
  constraints on its couplings from a combined ATLAS and CMS analysis of the
  LHC pp collision data at $ \sqrt{s}=7 $ and 8 TeV}.
\newblock {\em JHEP}, 08:045, 2016.

\bibitem{Fuentes-Martin:2019mun}
Javier Fuentes-Martín, Gino Isidori, Julie Pagès, and Kei Yamamoto.
\newblock {With or without U(2)? Probing non-standard flavor and helicity
  structures in semileptonic B decays}.
\newblock 2019.

\bibitem{Becirevic:2018afm}
Damir Bečirević, Ilja Doršner, Svjetlana Fajfer, Nejc Košnik, Darius~A.
  Faroughy, and Olcyr Sumensari.
\newblock {Scalar leptoquarks from grand unified theories to accommodate the
  $B$-physics anomalies}.
\newblock {\em Phys. Rev.}, D98(5):055003, 2018.

\bibitem{Pich:1998xt}
Antonio Pich.
\newblock {Effective field theory: Course}.
\newblock In {\em {Probing the standard model of particle interactions.
  Proceedings, Summer School in Theoretical Physics, NATO Advanced Study
  Institute, 68th session, Les Houches, France, July 28-September 5, 1997. Pt.
  1, 2}}, pages 949--1049, 1998.

\bibitem{Neubert:1996wg}
Matthias Neubert.
\newblock {Heavy quark effective theory}.
\newblock {\em Subnucl. Ser.}, 34:98--165, 1997.

\bibitem{Buras:1998raa}
Andrzej~J. Buras.
\newblock {Weak Hamiltonian, CP violation and rare decays}.
\newblock In {\em {Probing the standard model of particle interactions.
  Proceedings, Summer School in Theoretical Physics, NATO Advanced Study
  Institute, 68th session, Les Houches, France, July 28-September 5, 1997. Pt.
  1, 2}}, pages 281--539, 1998.

\bibitem{Silvestrini:2019sey}
Luca Silvestrini.
\newblock {Effective Theories for Quark Flavour Physics}.
\newblock In {\em {Les Houches summer school: EFT in Particle Physics and
  Cosmology Les Houches, Chamonix Valley, France, July 3-28, 2017}}, 2019.

\bibitem{PDG2018}
M.~Tanabashi et~al.
\newblock {\href{http://pdg.lbl.gov/}{Review of particle physics}}.
\newblock {\em Phys. Rev.}, D98:030001, 2018.

\bibitem{Colangelo:2000dp}
Pietro Colangelo and Alexander Khodjamirian.
\newblock {QCD sum rules, a modern perspective}.
\newblock pages 1495--1576, 2000.

\bibitem{Mathur:2002ce}
Nilmani Mathur, Randy Lewis, and R.M. Woloshyn.
\newblock {Charmed and bottom baryons from lattice NRQCD}.
\newblock {\em Phys. Rev.}, D66:014502, 2002.

\bibitem{Karliner:2009}
M.~Karliner, B.~Keren-Zur, H.J. Lipkin, and J.L. Rosner.
\newblock {The quark model and $b$ baryons}.
\newblock {\em Annals Phys.}, 342:2--15, 2009.

\bibitem{Jenkins:2007dm}
Elizabeth~Ellen Jenkins.
\newblock {Model-independent bottom baryon mass predictions in the $1/N_c$
  expansion}.
\newblock {\em Phys. Rev.}, D77:034012, 2008.

\bibitem{Ebert:2005xj}
D.~Ebert, R.N. Faustov, and V.O. Galkin.
\newblock {Masses of heavy baryons in the relativistic quark model}.
\newblock {\em Phys. Rev.}, D72:034026, 2005.

\bibitem{Liu:2007fg}
Xiang Liu, Hua-Xing Chen, Yan-Rui Liu, Atsushi Hosaka, and Shi-Lin Zhu.
\newblock {Bottom baryons}.
\newblock {\em Phys. Rev.}, D77:014031, 2008.

\bibitem{Zhang:2008rm}
Jian-Rong Zhang and Ming-Qiu Huang.
\newblock {Heavy baryon spectroscopy in QCD}.
\newblock {\em Phys. Rev.}, D78:094015, 2008.

\bibitem{Lewis:2009ce}
Randy Lewis and R.M. Woloshyn.
\newblock {Bottom baryons from a dynamical lattice QCD simulation}.
\newblock {\em Phys. Rev.}, D79:014502, 2009.

\bibitem{Lenz:2015dra}
Alexander Lenz.
\newblock {Lifetimes and heavy quark expansion}.
\newblock {\em Int. J. Mod. Phys.}, A30(10):1543005, 2015.
\newblock [,63(2014)].

\bibitem{Bigi:1995jr}
Ikaros I.~Y. Bigi.
\newblock {The QCD perspective on lifetimes of heavy flavor hadrons}.
\newblock 1995.

\bibitem{Cheng:1997xba}
Hai-Yang Cheng.
\newblock {A Phenomenological analysis of heavy hadron lifetimes}.
\newblock {\em Phys. Rev.}, D56:2783--2798, 1997.

\bibitem{Ito:1997qq}
Toshiaki Ito, Masahisa Matsuda, and Yoshimitsu Matsui.
\newblock {New possibility of solving the problem of lifetime ratio
  $\tau(\Lambda_b)/\tau(B_d)$}.
\newblock {\em Prog. Theor. Phys.}, 99:271--280, 1998.

\bibitem{Lattice}
C.~Aubin, C.~Bernard, C.~DeTar, M.~DiPierro, A.~El-Khadra, Steven Gottlieb,
  E.~B. Gregory, U.~M. Heller, J.~Hetrick, A.~S. Kronfeld, P.~B. Mackenzie,
  D.~Menscher, M.~Nobes, M.~Okamoto, M.~B. Oktay, J.~Osborn, J.~Simone,
  R.~Sugar, D.~Toussaint, and H.~D. Trottier.
\newblock Semileptonic decays of $d$ mesons in three-flavor lattice qcd.
\newblock {\em Phys. Rev. Lett.}, 94:011601, Jan 2005.

\bibitem{Wilson}
Christoph Bobeth, Mikolaj Misiak, and Jorg Urban.
\newblock {Photonic penguins at two loops and $m_t$ dependence of $BR[B \to X_s
  l^+ l^-]$}.
\newblock {\em Nucl. Phys.}, B574:291--330, 2000.

\bibitem{SMPredIsidori}
Marzia Bordone, Gino Isidori, and Andrea Pattori.
\newblock {On the Standard Model predictions for $R_K$ and $R_{K^*}$}.
\newblock {\em Eur. Phys. J.}, C76(8):440, 2016.

\bibitem{Amhis:2019ckw}
Yasmine~Sara Amhis et~al.
\newblock {Averages of $b$-hadron, $c$-hadron, and $\tau$-lepton properties as
  of 2018}.
\newblock 2019.

\bibitem{Aaij:2019wad}
Roel Aaij et~al.
\newblock {Search for lepton-universality violation in $B^+\to K^+\ell^+\ell^-$
  decays}.
\newblock {\em Phys. Rev. Lett.}, 122(19):191801, 2019.

\bibitem{Rkst}
R.~Aaij et~al.
\newblock {Test of lepton universality with $B^{0} \rightarrow
  K^{*0}\ell^{+}\ell^{-}$ decays}.
\newblock 2017.

\bibitem{Aaij:2015esa}
Roel Aaij et~al.
\newblock {Angular analysis and differential branching fraction of the decay
  $B^0_s\to\phi\mu^+\mu^-$}.
\newblock {\em JHEP}, 09:179, 2015.

\bibitem{Aaij:2014pli}
R.~Aaij et~al.
\newblock {Differential branching fractions and isospin asymmetries of $B \to
  K^{(*)} \mu^+ \mu^-$ decays}.
\newblock {\em JHEP}, 06:133, 2014.

\bibitem{Ciuchini:2018anp}
Marco Ciuchini, Antonio~M. Coutinho, Marco Fedele, Enrico Franco, Ayan Paul,
  Luca Silvestrini, and Mauro Valli.
\newblock {Hadronic uncertainties in semileptonic $B\to K^*\mu^+\mu^-$ decays}.
\newblock {\em PoS}, BEAUTY2018:044, 2018.

\bibitem{Aebischer:2019mlg}
Jason Aebischer, Wolfgang Altmannshofer, Diego Guadagnoli, Méril Reboud, Peter
  Stangl, and David~M. Straub.
\newblock {B-decay discrepancies after Moriond 2019}.
\newblock 2019.

\bibitem{DiLuzio:2017vat}
Luca Di~Luzio, Admir Greljo, and Marco Nardecchia.
\newblock {Gauge leptoquark as the origin of B-physics anomalies}.
\newblock {\em Phys. Rev.}, D96(11):115011, 2017.

\bibitem{Calibbi:2017qbu}
Lorenzo Calibbi, Andreas Crivellin, and Tianjun Li.
\newblock {Model of vector leptoquarks in view of the $B$-physics anomalies}.
\newblock {\em Phys. Rev.}, D98(11):115002, 2018.

\bibitem{Bordone:2017bld}
Marzia Bordone, Claudia Cornella, Javier Fuentes-Martin, and Gino Isidori.
\newblock {A three-site gauge model for flavor hierarchies and flavor
  anomalies}.
\newblock {\em Phys. Lett.}, B779:317--323, 2018.

\bibitem{Barbieri:2017tuq}
Riccardo Barbieri and Andrea Tesi.
\newblock {$B$-decay anomalies in Pati-Salam SU(4)}.
\newblock {\em Eur. Phys. J.}, C78(3):193, 2018.

\bibitem{Crivellin:2017zlb}
Andreas Crivellin, Dario Müller, and Toshihiko Ota.
\newblock {Simultaneous explanation of $R_{D^{(*)}}$ and $b\to s \mu^{+}
  \mu^{-}$: the last scalar leptoquarks standing}.
\newblock {\em JHEP}, 09:040, 2017.

\bibitem{Buttazzo:2017ixm}
Dario Buttazzo, Admir Greljo, Gino Isidori, and David Marzocca.
\newblock {B-physics anomalies: a guide to combined explanations}.
\newblock {\em JHEP}, 11:044, 2017.

\bibitem{Marzocca:2018}
David Marzocca.
\newblock {Addressing the B-physics anomalies in a fundamental Composite Higgs
  Model}.
\newblock {\em JHEP}, 07:121, 2018.

\bibitem{Angelescu:2018tyl}
A.~Angelescu, Damir Bečirević, D.~A. Faroughy, and O.~Sumensari.
\newblock {Closing the window on single leptoquark solutions to the $B$-physics
  anomalies}.
\newblock {\em JHEP}, 10:183, 2018.

\bibitem{Romao:2017qnu}
Miguel Crispim~Romao, Stephen~F. King, and George~K. Leontaris.
\newblock {Non-universal $Z'$ from fluxed GUTs}.
\newblock {\em Phys. Lett.}, B782:353--361, 2018.

\bibitem{BecirevicLQ}
Damir Becireciv and Olcyr Sumensari.
\newblock {A leptoquark model to accommodate $R_K^\mathrm{exp} <R_K^\mathrm{SM}
  $ and $R_{K^\ast}^\mathrm{exp}< R_{K^\ast}^\mathrm{SM}$}.
\newblock 2017.

\bibitem{Arnan:2019uhr}
Pere Arnan, Andreas Crivellin, Marco Fedele, and Federico Mescia.
\newblock {Generic loop effects of new scalars and fermions in $b\to
  s\ell^+\ell^-$ and a vector-like $4^{\rm th}$ generation}.
\newblock {\em JHEP}, 06:118, 2019.

\bibitem{LHCb}
Roel Aaij et~al.
\newblock {LHCb Detector Performance}.
\newblock {\em Int. J. Mod. Phys.}, A30(07):1530022, 2015.

\bibitem{Aaij:1978280}
{Aaij, Roel and others}.
\newblock {LHCb Detector Performance}.
\newblock {\em Int. J. Mod. Phys. A}, 30(LHCB-DP-2014-002. CERN-PH-EP-2014-290.
  LHCB-DP-2014-002. CERN-LHCB-DP-2014-002):1530022. 73 p, Dec 2014.

\bibitem{Bediaga:1443882}
I~Bediaga et~al.
\newblock {Framework TDR for the LHCb Upgrade: Technical Design Report}.
\newblock Technical Report CERN-LHCC-2012-007. LHCb-TDR-12, Apr 2012.

\bibitem{Collaboration:1624070}
LHCb Collaboration.
\newblock {LHCb VELO Upgrade Technical Design Report}.
\newblock Technical Report CERN-LHCC-2013-021. LHCB-TDR-013, Nov 2013.

\bibitem{Quagliani:2296404}
Renato Quagliani.
\newblock {Study of double charm B decays with the LHCb experiment at CERN and
  track reconstruction for the LHCb upgrade}, Oct 2017.
\newblock Presented 06 Oct 2017.

\bibitem{Collaboration:1647400}
LHCb Collaboration.
\newblock {LHCb Tracker Upgrade Technical Design Report}.
\newblock Technical Report CERN-LHCC-2014-001. LHCB-TDR-015, Feb 2014.

\bibitem{Amato:494264}
S~Amato et~al.
\newblock {\em {LHCb calorimeters: Technical Design Report}}.
\newblock Technical Design Report LHCb. CERN, Geneva, 2000.

\bibitem{Albajar:1991sq}
C.~Albajar et~al.
\newblock {First observation of the beauty baryon $\Lambda_b$ in the decay
  channel $\Lambda_b\to J/\psi\Lambda$ at the CERN proton - anti-proton
  collider}.
\newblock {\em Phys. Lett.}, B273:540--548, 1991.
\newblock [,249(1992)].

\bibitem{Aaij:2019pqz}
Roel Aaij et~al.
\newblock {Measurement of $b$ hadron fractions in 13 TeV $pp$ collisions}.
\newblock {\em Phys. Rev.}, D100:031102, 2019.

\bibitem{Aaij:2013qja}
R~Aaij et~al.
\newblock {Measurement of the $\Lambda_b^0$, $\Xi_b^-$ and $\Omega_b^-$ baryon
  masses}.
\newblock {\em Phys. Rev. Lett.}, 110(18):182001, 2013.

\bibitem{Karliner:2008sv}
Marek Karliner, Boaz Keren-Zur, Harry~J. Lipkin, and Jonathan~L. Rosner.
\newblock {The Quark Model and $b$ Baryons}.
\newblock {\em Annals Phys.}, 324:2--15, 2009.

\bibitem{Aad:2012shb}
Georges Aad et~al.
\newblock {Measurement of the \myLb lifetime and mass in the ATLAS experiment}.
\newblock 2012.

\bibitem{Aaltonen:2009ny}
T.~Aaltonen et~al.
\newblock {Observation of the \myOb baryon and measurement of the properties of
  the \myXb and \myOb baryons}.
\newblock {\em Phys. Rev.}, D80:072003, 2009.

\bibitem{Acosta:2005mq}
D.~Acosta et~al.
\newblock {Measurement of $b$ hadron masses in exclusive \jpsi decays with the
  CDF detector}.
\newblock {\em Phys. Rev. Lett.}, 96:202001, 2006.

\bibitem{Abazov:2008qm}
V.M. Abazov et~al.
\newblock {Observation of the doubly strange $b$ baryon \myOb}.
\newblock {\em Phys. Rev. Lett.}, 101:232002, 2008.

\bibitem{Abazov:2007am}
V.M. Abazov et~al.
\newblock {Direct observation of the strange $b$ baryon \myXb}.
\newblock {\em Phys. Rev. Lett.}, 99:052001, 2007.

\bibitem{PDG2012}
J.~Beringer et~al.
\newblock {\href{http://pdg.lbl.gov/}{Review of particle physics}}.
\newblock {\em Phys. Rev.}, D86:010001, 2012.

\bibitem{Aaij:2016dls}
Roel Aaij et~al.
\newblock {Measurement of the mass and lifetime of the $\Omega_b^-$ baryon}.
\newblock {\em Phys. Rev.}, D93(9):092007, 2016.

\bibitem{Aaij:2020cex}
Roel Aaij et~al.
\newblock {First observation of excited $\Omega_b^-$ states}.
\newblock {\em Phys. Rev. Lett.}, 124(8):082002, 2020.

\bibitem{Aaij:2014owa}
Roel Aaij et~al.
\newblock {Measurements of the $B^+, B^0, B^0_s$ meson and $\Lambda^0_b$ baryon
  lifetimes}.
\newblock {\em JHEP}, 04:114, 2014.

\bibitem{Pivk:2004ty}
Muriel Pivk and Francois~R. Le~Diberder.
\newblock {sPlot: a statistical tool to unfold data distributions}.
\newblock {\em Nucl.Instrum.Meth.}, A555:356--369, 2005.

\bibitem{Gabbiani:2004tp}
Fabrizio Gabbiani, Andrei~I. Onishchenko, and Alexey~A. Petrov.
\newblock {Spectator effects and lifetimes of heavy hadrons}.
\newblock {\em Phys. Rev.}, D70:094031, 2004.

\bibitem{Altarelli:1996gt}
Guido Altarelli, G.~Martinelli, S.~Petrarca, and F.~Rapuano.
\newblock {Failure of local duality in inclusive nonleptonic heavy flavor
  decays}.
\newblock {\em Phys. Lett.}, B382:409--414, 1996.

\bibitem{Gershon:2010wx}
Tim Gershon.
\newblock {$\Delta \Gamma_d$: a forgotten null test of the standard model}.
\newblock {\em J.Phys.}, G38:015007, 2011.

\bibitem{Lenz:2006hd}
Alexander Lenz and Ulrich Nierste.
\newblock {Theoretical update of $B_s - \bar{B}_s$ mixing}.
\newblock {\em JHEP}, 06:072, 2007.

\bibitem{Lenz:2011ti}
Alexander Lenz and Ulrich Nierste.
\newblock {Numerical updates of lifetimes and mixing parameters of B mesons}.
\newblock 2011.

\bibitem{Aaij:2019bzx}
Roel Aaij et~al.
\newblock {Test of lepton universality with $\Lambda^{0}_{b} \to p K^- \ell^+
  \ell^-$ decays}.
\newblock 2019.

\bibitem{Lisovskyi:2699822}
Vitalii Lisovskyi.
\newblock {Study of rare $b$-baryon decays and test of lepton universality at
  LHCb}, 2019.
\newblock Presented 09 Sep 2019.

\bibitem{ALEPH:2005ab}
S.~Schael et~al.
\newblock {Precision electroweak measurements on the $Z$ resonance}.
\newblock {\em Phys. Rept.}, 427:257--454, 2006.

\bibitem{Aaboud:2016btc}
Morad Aaboud et~al.
\newblock {Precision measurement and interpretation of inclusive $W^+$ , $W^-$
  and $Z/\gamma ^*$ production cross sections with the ATLAS detector}.
\newblock {\em Eur. Phys. J.}, C77(6):367, 2017.

\bibitem{Bifani:2018zmi}
Simone Bifani, Sébastien Descotes-Genon, Antonio Romero~Vidal, and
  Marie-Hélène Schune.
\newblock {Review of Lepton Universality tests in $B$ decays}.
\newblock {\em J. Phys.}, G46(2):023001, 2019.

\bibitem{LHCb-PAPER-2016-059}
R.~Aaij et~al.
\newblock {Observation of the decay $\Lb\to\proton\Km\mup\mun$ a search for
  \CP\ violation}.
\newblock {\em JHEP}, 06:1, 2017.

\bibitem{LHCb-PAPER-2015-032}
R.~Aaij et~al.
\newblock {Study of the productions of $\Lb$ and $\Bzb$ hadrons in
  $\proton\proton$ collisions and first measurement of the
  $\Lb\to\jpsi\proton\Km$ branching fraction}.
\newblock {\em Chin. Phys. C}, 40:011001, 2016.

\bibitem{LHCb-PAPER-2015-060}
R.~Aaij et~al.
\newblock {Observation of $\Lb\to\psitwos\proton\Km$ and
  $\Lb\to\jpsi\pip\pim\proton\Km$ decays and a measurement of the $\Lb$ baryon
  mass}.
\newblock {\em JHEP}, 05:132, 2016.

\bibitem{LHCb-PAPER-2015-029}
R.~Aaij et~al.
\newblock {Observation of $\jpsi\proton$ resonances consistent with pentaquark
  states in $\Lb\to\jpsi\proton\Km$ decays}.
\newblock {\em Phys. Rev. Lett.}, 115:072001, 2015.

\bibitem{Aaij:2636441}
Roel Aaij et~al.
\newblock {Physics case for an LHCb Upgrade II - Opportunities in flavour
  physics, and beyond, in the HL-LHC era}.
\newblock Technical Report LHCB-PUB-2018-009. LHCC-G-171, CERN, Geneva, Aug
  2018.
\newblock ISBN 978-92-9083-494-6.

\bibitem{scikit-learn}
F.~Pedregosa, G.~Varoquaux, A.~Gramfort, V.~Michel, B.~Thirion, O.~Grisel,
  M.~Blondel, P.~Prettenhofer, R.~Weiss, V.~Dubourg, J.~Vanderplas, A.~Passos,
  D.~Cournapeau, M.~Brucher, M.~Perrot, and E.~Duchesnay.
\newblock Scikit-learn: Machine learning in {P}ython.
\newblock {\em Journal of Machine Learning Research}, 12:2825--2830, 2011.

\bibitem{Straub:2018kue}
David~M. Straub.
\newblock {flavio: a python package for flavour and precision phenomenology in
  the Standard Model and beyond}.
\newblock 2018.

\end{thebibliography}

\end{document}